\documentclass[aps,preprint,superscriptaddress,amsmath,amssymb]{revtex4}
\usepackage{graphicx,color,slashed}
\usepackage{subfigure,bbold}

\def\be{\begin{eqnarray}}
\def\ed{\end{eqnarray}}
\def\non{\nonumber}

\begin{document}

{\begin{flushright}{KIAS-P15045}
\end{flushright}}

\title{ Diboson excess in the Higgs singlet and vector-like quark models  }

\author{ \bf Chuan-Hung Chen\footnote{Email:physchen@mail.ncku.edu.tw} }
\affiliation{Department of Physics, National Cheng-Kung
University, Tainan 70101, Taiwan  }

\author{\bf Takaaki Nomura\footnote{Email:nomura@kias.re.kr} }
\affiliation{School of Physics, Korea Institute for Advanced Study, Seoul 130-722, Republic of Korea}

\date{\today}

\begin{abstract}

Diboson resonance with mass  of $1.8-2$ TeV  is reported successively by  CMS and ATLAS experiments in proton-proton collisions at $\sqrt{s}=8$ TeV. We investigate the potentiality of Higgs singlet as the TeV resonance. The challenges of low production cross section and high width  for a fundamental scalar could be got over by three factors: (1) larger Yukawa couplings, (2) larger number of heavy quarks and (3) smaller mixing angle with standard model Higgs. We find that the required factors could be realized in the framework of two vector-like triplet quarks (VLTQs)  and  the  resulting production cross section and decay fraction  of heavy Higgs $\sigma(pp\to H )\times  {\rm BR}( H\to W^+ W^- + ZZ)$  can be of  ${\cal O}(10)$ fb when masses of new heavy quarks are 1 TeV, the values of Yukawa couplings are around $3$ and the mixing angle is  $\sin\theta \sim 0.11$. We also find that the SM Higgs production and its decay in the process $pp\to h \to \gamma \gamma$ could be still consistent with current data when  a color-triplet scalar $(3,3)_{1/3}$ is considered. Furthermore, we  study the  product of  VLTQ-pair production cross section  and the BRs of VLTQ  decays, and  find that the cross sections in the decay channels, such as $u_{4,5} \to b W^+$,  $d_5 \to t W^-$ and $d_4 \to b h(Z)$ could be  $7-17$ fb at 13 TeV LHC.

\end{abstract}

\maketitle

The searches of TeV diboson resonances  are performed in proton-proton collisions at $\sqrt{s}=8$ TeV  ATLAS~ \cite{Aad:2015owa} and CMS~\cite{Khachatryan:2014hpa} experiments.  Although a moderate diboson excess at around $1.8$ TeV is found by CMS~\cite{Khachatryan:2014hpa}  in the semileptonic channels,  however,  the diboson excess of $(WZ,\, WW,\, ZZ)$  observed by ATLAS  in the dijet invariant mass spectrum has the significance of $(3.4,\, 2.6,\, 2.9)\sigma$, respectively.

Analyzing the jet substructures and using the tagged jet mass $m_{j}$ determined by $|m_j -m_V|<13$ GeV, the reconstructed boson  from a single jet at ATLAS could be $W$ or $Z$ boson in the standard model (SM), and  the resulting  cross sections $\sigma(p p\to R)BR(R\to VV')$  are in the region of $16-30$ fb, where $R$ is the resonance and $V^{(\prime)}$ is the weak gauge boson $W/Z$.  In order to interpret the ATLAS excess, the possible candidates are 
a spin-2 Kaluza-Klein mode of the bulk Randall-Sundrum graviton~\cite{Aad:2015owa}, composite spin-1 particle~\cite{Fukano:2015hga, Carmona:2015xaa, Franzosi:2015zra,Bian:2015ota,Low:2015uha, Terazawa:2015bsa}, spin-1 bosons e.g. $W'$/$Z'$~\cite{Hisano:2015gna,Cheung:2015nha,Dobrescu:2015qna,Alves:2015mua, Gao:2015irw,Thamm:2015csa,Brehmer:2015cia,Cao:2015lia,Cacciapaglia:2015eea,Abe:2015jra,Abe:2015uaa, Dobrescu:2015yba, Anchordoqui:2015uea, Faraggi:2015iaa, Dev:2015pga, Goncalves:2015yua, Deppisch:2015cua}, composite spin-0 and/or spin-2 particles~\cite{Chiang:2015lqa,Cacciapaglia:2015nga, Sanz:2015zha}, scalar particles from extended Higgs sector and supersymmetric models~\cite{Chen:2015xql, Omura:2015nwa, Chao:2015eea, Petersson:2015rza, Zheng:2015dua}, and particles with effective interactions~\cite{Kim:2015vba, Arnan:2015csa, Fichet:2015yia}. A possible interpretation by triboson mode is also discussed in Ref.~\cite{Aguilar-Saavedra:2015rna}. 

In this work, we propose that the candidate of  diboson resonance with mass of around 2 TeV is the $SU(2)$ Higgs singlet  ($S$) which is predominantly produced by gluon-gluon fusion (ggF) $gg\to S$  and decays into $W^+ W^-/ZZ$ via the mixing with SM Higgs ($h$). The mechanism is motivated by  the following observation. It is known that the observed  scalar of 125 GeV in the SM is through the ggF production channel by top-quark loop effects and the effective interaction for $ggh$ could be written as \cite{Gunion:1989we}
 \be
 {\cal L}_{ggh} = \frac{\alpha_s}{12\pi} \frac{y_t}{\sqrt{2} m_t} N_F h G^a_{\mu \nu}  G^{a\mu\nu}\,, \label{eq:Lggh}
 \ed
where $y_t$ is the top-quark Yukawa coupling, the relation with vacuum expectation value (VEV) of SM Higgs is $m_t = y_t v /\sqrt{2}$, $N_F$ is the number of possible heavy quarks in the loop and $N_F=1$ in the SM. By the effective coupling, we see that the $h$ production cross section by ggF process could be enhanced by the Higgs Yukawa couplings to heavy quarks and  by the number of heavy quarks. For illustration, if we pretend $N_F= y_t=5$, $m_t=1$ TeV and $m_h =2$ TeV, the $h$ production cross section of ${\cal O}(10)$ fb can be achieved;  however, the cross section for $m_h=2$ TeV will be ${\cal O}(10^{-2})$ fb if  other values of parameters are not changed.  

In order to establish a model that obeys the SM gauge symmetry, owns a scalar with mass of around 2 TeV and naturally provides  larger $N_F$ and Yukawa couplings, we investigate the issue in the framework of vector-like quark (VLQ) model  with a heavy $SU(2)_L$ Higgs singlet. Basically, there is no limit for the  possible representations of VLQs. If we require  the VLQs those which can only mix with the SM up-type or down-type quarks, the possible representations are singlet, doublet and triplet~\cite{delAguila:2000rc, Okada:2012gy,Cacciapaglia:2012dd, Aguilar-Saavedra:2013qpa, deBlas:2014mba,Cacciapaglia:2015ixa}. For avoiding introducing too many VLQ states, we adopt the vector-like triplet quarks (VLTQs) in which each  triplet has three new quarks. In the base of gauge eigenstates, the introduced Higgs singlet  only couples to VLTQs and SM Higgs.  Therefore, if the masses of VLTQs and heavy Higgs are comparable, the main decay channels of the heavy scalar will be $gg$, $hh$ and $W^+ W^-/ZZ$ and the resulting total width could be below ${\cal O}(100)$ GeV, which can match the condition of  narrow resonance observed at the LHC.  Although the new heavy quarks could also enhance the SM Higgs production, however the enhancement indeed could be smeared when  colored scalar particles are introduced. Below, we discuss the model and its implications at the LHC.

We start to setup the model. In order to possess a heavy boson and several heavy quarks naturally, we extend the SM by including one real Higgs singlet and two VLTQs, where the representations of VLTQs in $SU(3)_c\times SU(2)_L \times U(1)_Y$ gauge symmetry are chosen as $(3,3)_{2/3}$ and $(3,3)_{-1/3}$ \cite{Okada:2012gy}.  For suppressing the new effects on the SM Higgs production cross section, we also add one color-triplet and $SU(2)$-triplet scalar $(3,3)_{1/3}$ to the model. In order to discuss the couplings of scalars to fermions, we first analyze the new scalar potential and write the  gauge invariant form to be 
 \begin{align}
 V(H,S) &= \mu^2_1 H^\dagger H + \lambda_1 (H^\dagger H)^2 + m^2_S S^2 + \mu_2 S^3  + 
 \mu_3 S (H^\dagger H) +  \lambda_2 S^4 + \lambda_3 S^2 (H^\dagger H) \non \\
 &+ m^2_\Psi \Psi^\dagger \Psi + \mu_4 \Psi^\dagger \Psi S+ \lambda_4 (\Psi^\dagger \Psi)^2+ \lambda_5 \Psi^\dagger\Psi H^\dagger H 
 + \lambda_6 \Psi^\dagger \Psi S^2\,. \label{eq:VHS}
 \end{align}
The representations of SM Higgs doublet, Higgs singlet and color-triplet  are taken by
 \be
H= \left(\begin{array}{cc}
 G^+    \\
\frac{1}{\sqrt{2}} ( v+ \phi + iG^0)     
\end{array}
  \right)\,, \ \ S = \frac{1}{\sqrt{2}} ( v_s + \Phi)\,, \ \ \Psi_\alpha = \left(\begin{array}{ccc}
 \Psi^{4/3}    \\
\Psi^{1/3} \\
 \Psi^{-2/3} \\
\end{array}
  \right)_\alpha\,, 
\ed
where $G^+$ and $G^0$ are Goldstone bosons, $\phi$ is the SM Higgs field and $v (v_s)$ is the VEV of $H$ ($S$). 
In our approach, the singlet $S$ has been a massive particle before electroweak symmetry breaking; therefore, basically a nonzero  VEV of S is not necessary, however Eq.~(\ref{eq:VHS}) could still lead to a nonzero $v_s$. By minimal conditions $\partial V(v,v_s)/\partial v =0$ and $\partial V(v,v_s)/\partial v_s =0$, we get 
 \begin{align}
 &\mu^2_1 + \lambda_1 v^2 + \frac{\mu_3 v_s}{\sqrt{2}} + \frac{\lambda_3 v^2_s}{2} =0\,, \non \\
 &m^2_S v_s + \frac{3 \mu_2 v^2_s}{2\sqrt{2}} + \frac{\mu_3 v^2}{2\sqrt{2}} + \lambda_2 v^3_s + \frac{\lambda_3 v_s v^2}{2} =0\,.
 \end{align}
 If we adopt $m^2_S \gg \mu_2 v_s, v^2_s$, the leading VEV of $S$ could be simplified by $v_s \approx - \mu_3 v^2 /(2\sqrt{2} m^2_S)$.  Therefore, when $m_H=2$ TeV, even $\mu_3 \sim m_S$, we still have $v_s \ll v$ .
  Using the scalar potential of Eq.~(\ref{eq:VHS}), the mass square matrix of $\phi$ and $\Phi$ is found by
 \be
\left(
\begin{array}{cc}
 \phi,  &  \Phi     
\end{array}
\right) 
\left(
\begin{array}{cc}
 m^2_\phi  &  \mu_3 v/\sqrt{2} + \lambda_3 v v_s \\
\mu_3 v/\sqrt{2} + \lambda_3 v v_s   & m^2_S + \lambda_3 v^2  /2   
\end{array}
\right) 
\left(
\begin{array}{c}
 \phi \\
 \Phi
\end{array}
\right) \label{eq:m2}
 \ed
with $m_\phi = \sqrt{2 \lambda_1} v$. The  parameters $\mu_3$ and $\lambda_3$ lead to the mixture of $\phi$ and $\Phi$. Since two scalar bosons are involved in the model, we only need one mixing angle to parametrize the mixing effect. As usual, we formulate the mass eigenstates to be
 \be
\left(
\begin{array}{c}
 h \\
 H
\end{array}
\right) = 
\left(
\begin{array}{cc}
\cos\theta  & -\sin\theta \\
\sin\theta  & \cos\theta    
\end{array}
\right) 
\left(
\begin{array}{c}
 \phi \\
 \Phi
\end{array}
\right)\,, \label{eq:mixing}
 \ed
where $h$ is the SM-like Higgs boson, $H$ is the new heavy Higgs boson and  the candidate of new resonance, and their masses are obtained as
 \be
 m^2_{h(H)} =  \frac{1}{2} \left[ \left( m^2_\phi + m^{\prime 2}_S \right) \mp \left( (m^{\prime 2}_S -m^2_\phi)^2 + 4 m^4_{\phi\Phi} \right)^{1/2} \right]
 \label{eq:Hmass}
 \ed
with $m^{\prime 2}_S = m^2_S + \lambda_3 v^2/2$ and $m^2_{\phi\Phi}= \mu_3 v/\sqrt{2} + \lambda_3 v v_s$.  The relationship of $m^2_{\phi\Phi}$ and mixing angle  can be expressed by $\sin2\theta = 2 m^2_{\phi \Phi}/(m^2_H -m^2_h)$. 

 The gauge invariant Yukawa couplings of VLTQs to the SM quarks, to the SM Higgs doublet and to the new Higgs singlet are written as 
\be
-{\cal L}^{Y}_{\rm VLTQ} &=&  \bar Q_L {\bf Y_1} F_{1R} \tilde{H}  + \bar Q_L {\bf Y_2} F_{2R} H+  \tilde y_{1}  Tr(\bar F_{1L}  F_{1R} ) S
+ \tilde y_2  Tr(\bar F_{2L} F_{2R} ) S \non \\
&+& M_{F_1} Tr(\bar F_{1L}  F_{1R} ) + M_{F_2} Tr(\bar F_{2L}  F_{2R})+ h.c.\,,  \label{eq:yukawa}
\ed
where $Q_L$ is the left-handed SM quark doublet and it could be regarded as mass eigenstate before VLTQs are introduced, all flavor indices are hidden,  $\tilde H =i \tau_2 H^*$,  $F_{1(2)}$ is the $2\times 2$ VLTQ with hypercharge $2/3(-1/3)$ and the representations of  $F_{1,2}$ in $SU(2)_L$ are expressed by
  \be
F_{1} = 
\left(
\begin{array}{cc}
 U_1/\sqrt{2} & X    \\
 D_1 &  -U_1/\sqrt{2}     
\end{array}
\right)\,, \  F_{2} = 
\left(
\begin{array}{cc}
 D_2/\sqrt{2} & U_2    \\
 Y &  -D_2/\sqrt{2}     
\end{array}
\right)\,.
\ed
The electric charges of $U_{1,2}$, $D_{1,2}$, $X$ and $Y$ are $2/3$, $-1/3$, $5/3$ and $-4/3$, respectively. Therefore, $U_{1,2} (D_{1,2})$ could mix with up (down) type SM quarks. $M_{F_{1(2)}}$ is the  mass of VLTQ, and due to the gauge symmetry, the VLTQs in the same multiplet state are degenerate.  By the Yukawa couplings of Eq.~(\ref{eq:yukawa}), the $5\times 5$ mass matrices for up and down type quarks are found by
\be
M_u = \left(
\begin{array}{ccc}
\left({\bf m}^{\rm dia}_{u} \right)_{3\times 3} |& v {\bf Y}_1/2  &  v {\bf Y}_2/\sqrt{2}  \\ 
 ----\ |& ---- &---- \\
\hspace{0.5cm} {\bf 0}_{2\times 3} \hspace{0.6 cm} |&    & \hspace{-1.2 cm}\left({\bf m}_{F}\right)_{2\times 2}
\end{array}
\right)\,, \  M_d = \left(
\begin{array}{ccc}
\left({\bf m}^{\rm dia}_{d} \right)_{3\times 3} |&  v {\bf Y}_1/\sqrt{2}  & -v {\bf Y}_2 / 2  \\ 
----\ | & ---- & ---- \\
\hspace{0.5cm}{\bf 0}_{2\times 3} \hspace{0.6cm}|&   & \hspace{-1.2cm} \left({\bf m}_{F}\right)_{2\times 2}
\end{array}
\right)\,, \label{eq:mass}
\ed
where $({\bf m}^{\rm dia}_{u})_{3\times 3}$ and $({\bf m}^{\rm dia}_{d})_{3\times 3}$ denote the  diagonal mass matrices  of SM quarks and ${\rm dia}({\bf m}_{F})_{2\times 2} = (m_{F_1}, m_{F_2})$. We note that a non-vanished $v_s$ could shift the masses of VLTQs. Since $v_s \ll v$, hereafter we neglect the small effects. Due to the presence of ${\bf Y_{1,2}}$, the SM quarks, $U_{1,2}$ and $D_{1,2}$ are not physical states anymore;  thus one has to diagonalize $M_u$ and $M_d$ to get the mass eigenstates. If $v Y^i_{1,2} \ll m_{F_{1,2}}$, we expect that the off-diagonal elements of unitary matrices for diagonalizing the mass matrices should be  of order of $ v Y^i_{1,2}/m_{F_{1,2}}$. By adjusting  $Y^i_{1,2}$,  the off-diagonal effects could be enhanced and lead to interesting phenomena in collider physics.   

Besides the flavor conserving couplings, the Yukawa interactions in Eq.~(\ref{eq:yukawa}) also provide $\phi$- and $\Phi$-mediated flavor changing neutral currents (FCNCs) at the tree level.  Hence, the couplings of $\phi$ and $\Phi$ to quarks are written as
\be
-{\cal L}_{\phi (S)qq} &=& \frac{\phi}{v} \left[ \bar u_{L} M^{\rm dia}_{u} u_R + \bar d_{L} M^{\rm dia}_{d} d_R  \right] - \frac{\phi}{v}\left[ \bar u_L V^u_L {\cal M}_U V^{u\dagger}_R u_R + \bar d_L V^d_L {\cal M}_D V^{d\dagger}_R d_R \right] \non \\
&+&  \Phi  \left[ \bar u_L V^u_L {\cal Y} V^{u\dagger}_{R} u_R + \bar d_L V^d_L {\cal Y} V^{d\dagger}_{R} d_R\right]+h.c. \,, \label{eq:hqq}
\ed
where $u$ and $d$ stand for the five up and down type quarks in flavor space and the flavor indices are not shown explicitly, $V^q_L$ and $V^q_R$ are the unitary matrices for diagonalizing the mass matrix defined in Eq.~(\ref{eq:mass}), $M^{\rm dia}_{q} = V^q_L M_{q} V^{q\dagger}_R$, dia${\cal M}_F={\rm dia}{\cal M}_{U,D}=(0,0,0, m_{F_1}, m_{F_2})$ and dia${\cal Y}=(0,0,0, y_1, y_2)$ with $y_{1, 2} = \tilde y_{1, 2}/\sqrt{2}$. The first brackets in Eq.~(\ref{eq:hqq}) only give the flavor conserving couplings while the tree level FCNCs are from the second and third brackets. The Yukawa couplings of $h$ and $H$ could be easily obtained by using Eq.~(\ref{eq:mixing}). In general, although the off-diagonal elements of $V^q_{L,R}$ are free parameters,  one can use $M_q M^{\dagger}_q$ and $M^{\dagger}_q M_q$ to get more useful information, where the former is only associated with  $V^q_L$ and the latter is $V^q_R$. Due to the mass structures of Eq.~(\ref{eq:mass}), we further find that the off-diagonal elements in $M_q M^{\dagger}_q$ are proportional to $v {\bf Y}_i m_{F_i}$ while those elements in $M^{\dagger}_q M_q$ are ${\bf m}^{\rm dia}_{q} {\bf Y}_i v$. If we neglect the effects of $m_q Y_i v/ m^2_{F_j}$, it will be a good approximation to set $V^q_R \approx 1$. That is, the tree level FCNCs are predominantly arisen from $V^q_L$. 

Although the FCNC effects do not affect the production cross section of new resonance, however, the search of VLQ at colliders depends on the couplings. In order to study the signals at the LHC, here we discuss the simple scheme for $V^q_L$ based on the viewpoint of phenomenological analysis. When the mass matrix $M_q$ is diagonalized, the masses of light quarks should be maintained. Therefore, we require that VLTQs only couple to third generation of SM quark, i.e. $Y_{11} = Y_{12}=Y_{21}=Y_{22}=0$. Consequently, the mass matrices in Eq.~(\ref{eq:mass}) can be reduced to $3\times 3$ matrices and are written as
 \be
 {\cal M}_u = \left(
\begin{array}{ccc}
m_t & v Y_{13}/2  &  v Y_{23}/\sqrt{2}  \\ 
0&  m_{F_1}  & 0  \\
0 & 0 & m_{F_2} 
\end{array}
\right)\,, \  \ \  {\cal M}_d = \left(
\begin{array}{ccc}
m_b & v Y_{13}/\sqrt{2}  &  -v Y_{23}/2  \\ 
0&  m_{F_1}  & 0  \\
0 & 0 & m_{F_2} 
\end{array}
\right)\,.
 \ed 
 If $\zeta_i= vY_{i3}/m_{F_i}\ll 1$ is satisfied and $\zeta^2_i$ is dropped, we find that the associated matrix ${\cal V}^q_L$ for diagonalizing ${\cal M}_q {\cal M}^\dagger_{q}$ can be parametrized by
  \be
  {\cal V}^u_L \approx \left(
\begin{array}{ccc}
1 & -\zeta_1/2  &  -\zeta_2/\sqrt{2}  \\ 
 \zeta_1/2  &  1  & 0  \\
 \zeta_2/\sqrt{2}   & 0 & 1
\end{array}
\right)\,, \ \    {\cal V}^d_L \approx \left(
\begin{array}{ccc}
1 & -\zeta_1/\sqrt{2}  &  \zeta_2/2  \\ 
 \zeta_1/\sqrt{2}  &  1  & 0  \\
-\zeta_2/2   & 0 & 1
\end{array}
\right) \,. \label{eq:VqL}
  \ed
We note that  the unitary matrices in Eq.~(\ref{eq:VqL}) are constructed to diagonalize ${\cal M}_q {\cal M}^\dagger_{q}$. Under the approximations of $\zeta_i \zeta_j \approx 0$ and $V^q_R \approx 1$ in which $m_{t,b} v Y_{i3}/m^2_{F_i}$ are neglected, the same factors appearing in the off-diagonal elements of ${\cal M}^{\rm dia}_q = {\cal V}^q_{L} {\cal M}_q {\cal V}^q_R \approx {\cal V}^q_{L} {\cal M}_q $ should be also dropped. 
 In order to get the mixing matrix $V^q_L$ in five flavors, we can set $V^{q}_{L11}=V^{q}_{L22}=1$, $V^{q}_{L1k}=V^{q}_{L2m}=0$ with $k=2\sim 5$ and $m=1,3\sim 5$ and $V^q_{L \alpha \beta}={\cal V}^q_{Lij}$ with $\alpha=i+2$, $\beta=j+2$ and $i,j=1\sim 3$.  Using Eqs.~(\ref{eq:mixing}) and (\ref{eq:VqL}), the Higgs-mediated FCNCs associated with $t$ and $b$ quarks are found as 
\be
{\cal L}_{hQq} &=&\left(  \cos\theta \frac{{\cal M}_{FJJ}}{v} + \sin\theta {\cal Y}_J  \right) \left[  (V^u_L)_{3J} \bar t_L u_{JR} + (V^d_L)_{3J}  \bar b_L d_{JR} \right] h \non \\
&-&  \left(  \cos\theta {\cal Y}_J  - \sin\theta \frac{{\cal M}_{FJJ}}{v}  \right) \left[ (V^u_L)_{3J} \bar t_L u_{JR} + (V^d_L)_{3J} \bar b_L d_{JR} \right] H+ h.c.\,, \label{eq:hHQq}
\ed
where   $J=4,5$ stand for the new heavy quarks with electric charge of $2/3$ or $-1/3$.

Next, we discuss the weak interactions of VLTQs.  As usual, we write the covariant derivative of $SU(2)_L\times U(1)_Y$ as 
  \be
 D_\mu = \partial_\mu + i \frac{g}{\sqrt{2}} \left( T^+ W^+_\mu +T^- W^{-}_\mu \right) + i\frac{g}{c_W} \left(   T_3 -  s^2_W Q \right) Z_\mu+ i e Q A_\mu \,,
 \ed
where $W^{\pm}_\mu$, $Z_\mu$ and $A_\mu$ stand for the gauge bosons in the SM, $g$ is the gauge coupling of $SU(2)_L$, $s_W(c_W)=\sin\theta_W (\cos\theta_W)$, $\theta_W$ is the Weinberg angle, $T^{\pm} =T_1 \pm i T_2$ and the charge operator $Q = T_3 + Y$ with $Y$ being the hypercharge of particle.  The generators of $SU(2)$ in triplet representation are set to be 
 \be
  T_1=\frac{1}{\sqrt{2}}\begin{pmatrix}
  0 & 1 & 0 \\
   1 & 0 & 1 \\
    0 & 1 & 0 \\
  \end{pmatrix}\,,
~ T_2= \frac{1}{\sqrt{2}} \begin{pmatrix}     0 & -i & 0 \\
   i & 0 & -i \\
    0 & i & 0 \\
     \end{pmatrix}\,,~  T_3= \begin{pmatrix}
   1  & 0 & 0 \\
   0  & 0 &  0 \\
   0  & 0 & -1 \\
     \end{pmatrix}\,. 
 \ed
Accordingly, the gauge interactions of new quarks are summarized by
 \be
 {\cal L}_{GFF} &=& -g \left[\left( \bar X \gamma^\mu U_1 + \bar U_1 \gamma^\mu D_1 + \bar D_2 \gamma^\mu Y + \bar U_2 \gamma^\mu D_2 \right) W^+_\mu + h.c. \right] \non \\
 &-& \left[ \frac{g}{c_W}  \bar F_1 \left( T^3 -s^2_W Q_1 \right) F_1 Z_\mu + e \bar F_1 \gamma^\mu Q_1 F_1 A_\mu  + ( F_1 \to F_2, Q_1 \to Q_2 ) \right]\,, 
 \ed
where the VLTQs  should be read by  $F^T_1 = (X, U_1, D_1)$ and $F^T_2 = (U_2, D_2, Y)$ and the associated charge operators are dia$Q_1= (5/3, 2/3, -1/3)$ and dia$Q_2 = (2/3, -1/3, -4/3)$.  As a result, the charged current interactions are written by
 \be
 {\cal L}_{Wud} &=& -\frac{g}{\sqrt{2}} \bar u_L \gamma^\mu V^L_{\rm CKM} d_L W^+_\mu 
  -\frac{g}{\sqrt{2}} \bar u_R \gamma^\mu V^R_{\rm CKM} d_R W^+_\mu +h.c.\,, 
 \ed
where $u$ and $d$ are the up and down type quarks in physical states, $V^{L(R)}_{\rm CKM}$ is the $5\times 5$ Cabibbo-Kobayashi-Maskawa (CKM) matrix for left(right)-handed quarks and their expressions are given by
 \be
 V^L_{\rm CKM} &=& V^u_L \left(
\begin{array}{cc}
\left(V_{\rm CKM} \right)_{3\times 3} |& {\bf 0}_{3\times 2}  \\ 
 ----\ \; |& ----  \\
\hspace{0.5cm} {\bf 0}_{2\times 3} \hspace{0.65 cm} |&  \sqrt{2} \mathbb{1}_{2\times 2}  
\end{array}
\right) %
 V^{d\dagger}_L \,, \ \  V^R_{\rm CKM} =  V^u_R 
  \left(
\begin{array}{cc}
{\bf 0}_{3\times 3} \ \ \, |& {\bf 0}_{3\times 2}  \\ 
 ---  |& --- \\
 \hspace{0.2cm}{\bf 0}_{2\times 3} \hspace{0.2cm}|&  \sqrt{2} \mathbb{1}_{2\times 2}  
\end{array}
\right)
 V^{d\dagger}_{R}\,.
 \ed
Here $(V_{\rm CKM})_{3\times 3}$ is the SM CKM matrix without VLTQs. Since the weak isospin of triplet quark differs from doublet quark, the new CKM matrices are not unitary matrices. Taking the approximation of $\zeta^2_i \approx 0$, the couplings of VLTQs to the third generation quarks are found by
 \be
 {\cal L}_{WQq} &=&- \frac{g}{\sqrt{2} } \left[ -\frac{3}{2} \zeta_2 \delta_{J5} \bar t_{L} \gamma^\mu d_{JL}  + \left( -\frac{\zeta_1}{2} \delta_{J4} +\sqrt{2} \zeta_2 \delta_{J5} \right)\bar u_{JL} \gamma^\mu b_L \right] W^+_\mu  +h.c.
 \label{eq:WQq}
 \ed 
with $J=4,5$. 

 By algebraic calculations, the weak neutral current interactions could be grouped to be 
 \be
 {\cal L}_{Zqq} &=& -\frac{g}{c_W} C^{q_L}_{ij} \bar q_{iL} \gamma^\mu q_{j L} Z_\mu - \frac{g}{c_W} C^{q_R}_{ij} \bar q_{i R} \gamma^\mu q_{jR} Z_\mu  
 \ed
and 
  \be
  C^{q_L}_{ij} &=&( I_3 -s^2_W Q_q )\delta_{ij} + \frac{1}{2} \left( -V^q_{Li4} V^{q*}_{Lj4} + V^q_{Li5} V^{q*}_{Lj5} \right)\,, \non \\
  C^{q_R}_{ij} &=&-s^2_W Q_q \delta_{ij} + \epsilon_q (V^q_{R})_{i \alpha_q} (V^{q*}_{R})_{\alpha_q j}
  \ed
with $I_{3}=\pm 1/2$ for up(down)-type quark, $(\epsilon_u, \alpha_u)= (1, 5)$ and $(\epsilon_d, \alpha_d)= (-1, 4)$. The second terms in $C^{q_L}_{ij}$ and $C^{q_R}_{ij}$ cause the tree level Z-mediated FCNCs. Using the results in Eq.~(\ref{eq:VqL}), the Z-mediated FCNCs associated with $t$ and $b$ quarks are given by
\be
 {\cal L}_{ZQq} &=& -\frac{g}{c_W}  \left( c^u_J \bar u_{JL} \gamma^\mu t_L +  c^d_J\bar d_{JL} \gamma^\mu b_L \right) Z_\mu + h.c.  
 \label{eq:ZQq}
 \ed
with $c^u_4 = \zeta_1/4$, $c^u_5 = -\zeta_2/2\sqrt{2}$, $c^d_4= \zeta_1/2\sqrt{2}$ and $c^d_5=\zeta_2/4$.

After introducing the model, we analyze the production of $H$ and its decays at $8$ TeV LHC. Since $H$ mainly couples to VLTQs, its production is through one-loop ggF processes. Due to the mixture of $h$ and $H$, $ggH$ effective coupling could be also induced by the color-triplet states. Thus,  the loop induced effective coupling for $ggH$ from VLTQs and $\Psi_\alpha$ is written by
  \be
{\cal L}_{ggH} = \frac{\alpha_s}{8\pi v} \left( \sum_{i=1,2} \frac{N_{F_i} y_i v}{2 m_{F_i}} A_{1/2}(\tau_i) \cos \theta  -\frac{ N_\Psi \lambda_5 v^2}{ m^2_\Psi} C(3) A_0 (\xi) \sin \theta \right) H G^{a\mu \nu}G^a_{\mu \nu} \,,\label{eq:LggH}
 \ed
where we have set $\mu_4$ in Eq.~(\ref{eq:VHS}) to be small, $N_{F_i}= 3$ is the number of  VLTQ in $F_{1(2)}$, $N_\Psi=3$ is the number of colored scalars, $C(3)=1/2$ is from the color factor  of color triplet $\Psi_\alpha$ and the loop functions are
 \begin{align}
A_{1/2}(\tau) &=  2 \tau [1+(1-\tau) f(\tau)^2]\,, \non  \\
A_0(x) &= x (1 -x f(x)^2 )
  \end{align}
  with $\tau_i = 4 m_{F_i}^2/m_H^2$, $\xi= 4 m^2_{\Psi}/m^2_H$ and $f(x)=\sin^{-1}(1/\sqrt{x})$. The same effects could also generate effective coupling $ggh$ and its expression is given by
  \be
  {\cal L}_{ggh} = \frac{\alpha_s}{8\pi v} \left( \sum_{i=1,2} \frac{N_{F_i} y_i v}{2m_{F_i}} A_{1/2}(\tau_{ih}) \sin \theta  -\frac{N_\Psi \lambda_5 v^2}{ m^2_\Psi} C(3) A_0 (\xi_h) \cos \theta \right) h G^{a\mu \nu}G^a_{\mu \nu}
  \ed
with $\tau_{ih}=4 m^2_{F_i}/m^2_h$ and $\xi_h=4 m^2_\Psi/m^2_h$. If we adopt 
 \be
 \sum_{i=1,2} \frac{N_{F_i} y_i v}{2m_{F_i}} A_{1/2}(\tau_{ih}) \sin \theta  \sim \frac{ N_\Psi \lambda_5 v^2}{ m^2_\Psi} C(3) A_0 (\xi_h) \cos \theta\,, \label{eq:con}
 \ed
the new effects on $ggh$ could be suppressed;  on the other hand, the contribution of  color-triplet scalar to $gg H$ is proportional to $\sin^2\theta/\cos\theta$ and it could be neglected in the scheme of small mixing angle.  Using the condition of Eq.~(\ref{eq:con}), the ratio to the SM result in $\Gamma(h\to \gamma\gamma)$ could be derived by \cite{Muhlleitner:2006wx}
 \be
R_{h\to \gamma\gamma}= \left| 1 + \frac{N_c y v  \sin\theta A_{1/2}(\tau_h)}{m_F} \frac{\sum_{F_i} Q^2_{F_i} - 2 \sum_{\Psi_i} Q^2_{\Psi_i} }{A_1( x_W) + 4/3 A_{1/2}(x_t) }\right|^2\,.
 \ed
With $Q(\Psi_\alpha )= ( 4/3, 1/3, -2/3)$ and the charges of VLTQs, we find $\sum_{F_i} Q^2_{F_i} - 2 \sum_{\Psi_i} Q^2_{\Psi_2} =1$. If we take $y=3$, $\sin\theta =0.08$ and $m_F=1$ TeV, we get $\mu_{h\to \gamma \gamma} \approx 0.93$. That is, the ratio of cross section to the SM prediction  for $pp\to h \to \gamma\gamma$  in our model is $\sigma(pp\to h)/\sigma(pp\to h)_{\rm SM} BR(h\to \gamma\gamma)/BR_{\rm SM}(h\to \gamma\gamma) \approx 0.93$, where the measurements by ATLAS and CMS are $1.17\pm 0.27$~\cite{Aad:2015gba} and $1.13\pm 0.24$~\cite{CMS}, respectively.
Hence, in our model, the product of $h$ production cross section and $BR(h\to \gamma\gamma)$ could be consistent with the LHC data, while the $H$ production is still enhanced by VLTQs.

\begin{figure}[hptb] 
\begin{center}
\includegraphics[width=70mm]{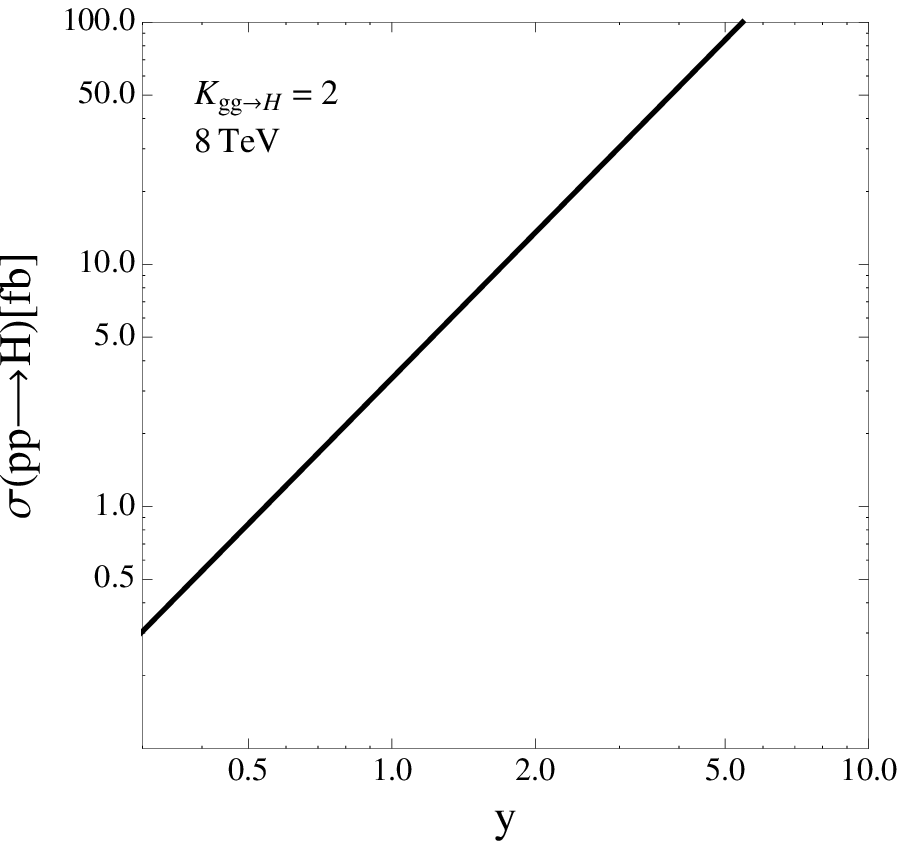} 
\includegraphics[width=70mm]{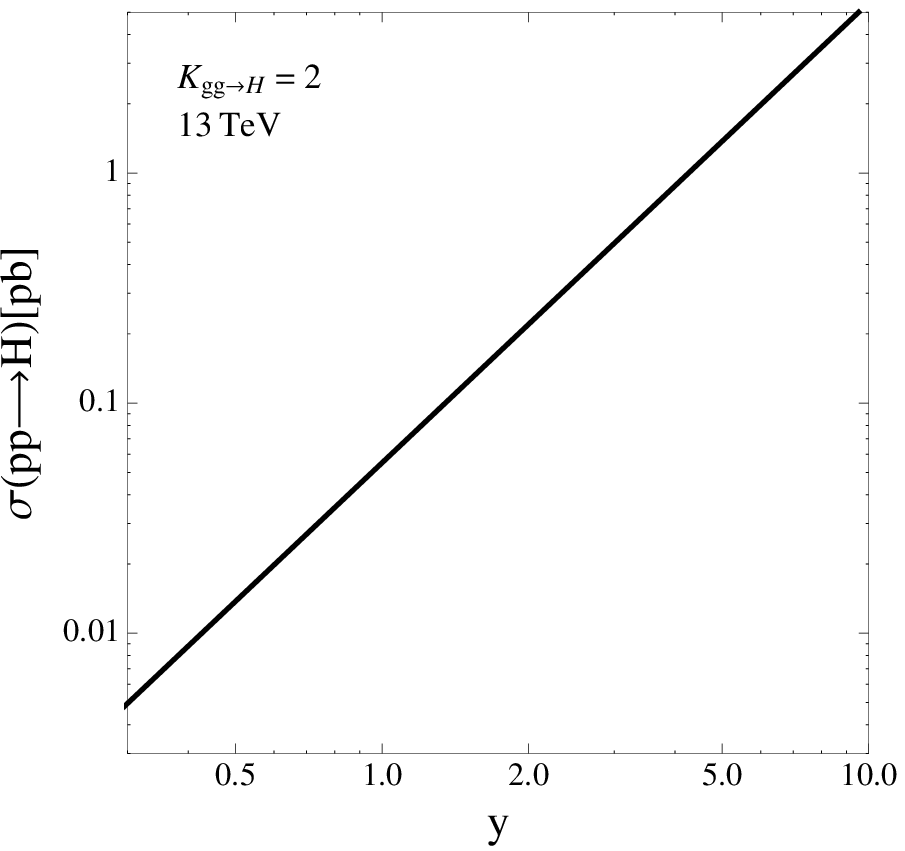} 
\caption{ Left: cross section (in units of fb)for $pp \to H$ process as a function of $y$ at $\sqrt{s}=8$ TeV, where we have used $m_F=m_{F_1}=m_{F_2}=1$ TeV and set $y=y_1=y_2$.  Right: $\sigma(pp\to H)$ (in units of pb) at 13 TeV. }
\label{fig:XS}
\end{center}
\end{figure}

  For numerically estimating the $H$ production cross section,  we implement the effective interaction of Eq.~(\ref{eq:LggH}) to CalcHEP~\cite{Belyaev:2012qa} and use  {\tt CTEQ6L} PDF~\cite{Nadolsky:2008zw}.  With $m_F=m_{F_1}=m_{F_2}=1$ TeV, $y=y_1=y_2$ and $\sqrt{s}=8$ TeV, the cross section for $pp\to H$ as a function of $y$ is presented in the left panel of Fig.~\ref{fig:XS}. For including the next-leading-order (NLO) effects, we have used the K-factor of $K_{gg\to H}=2.0$~\cite{Djouadi:2005gi}.  The factorization and renormalization scales are chosen to be $\mu=m_H/2$.  As a result, the production cross section of ${\cal O}(20)$ fb can be achieved while $m_F=1$ TeV and $y\sim 3$. For comparison, we also show $\sigma(pp\to H)$ at $\sqrt{s}=13$ TeV for LHC run 2. It is clear that the production cross section will be one order of magnitude larger than that from run 1.
%

\begin{figure}[h!] 
\begin{center}
\includegraphics[width=70mm]{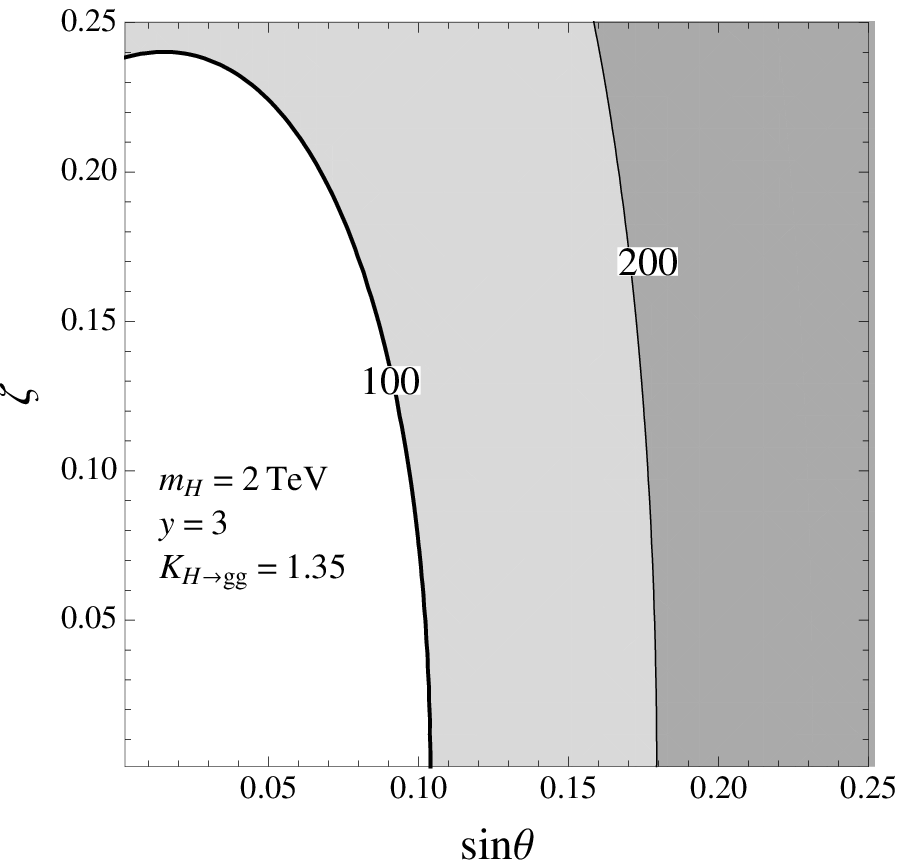} 
\includegraphics[width=70mm]{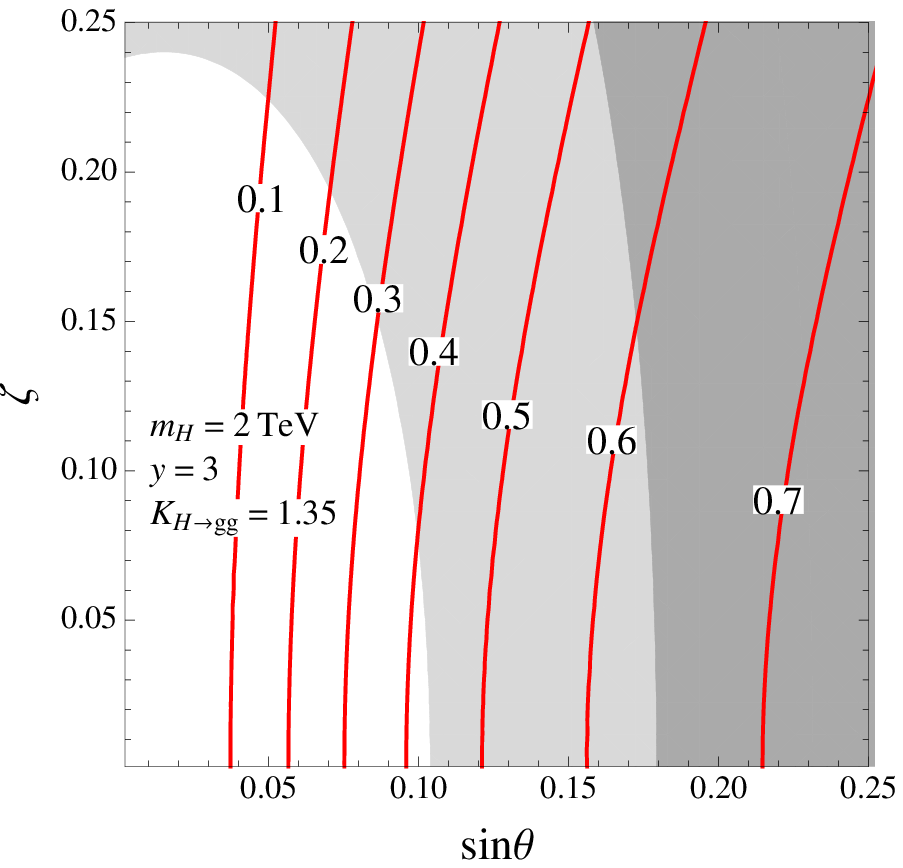} 
\caption{ Contours  for $\Gamma_{H}$ ( in units of GeV) [left] and for branching ratio of $H\to W^+W^- + ZZ$ [right] as a function of $\sin \theta$ and $\zeta$, where we have set $\zeta=\zeta_1=\zeta_2$ and  taken $(m_{H}, m_F)=(2, 1)$ TeV and $y = 3$, and the K-factor is $K_{H\to gg}=1.35$.  }
\label{fig:Gamma}
\end{center}
\end{figure}

Next, we discuss the decays of $H$. Based on Eq.~(\ref{eq:hqq}),  $H$ couplings to SM quarks are suppressed by $\zeta^2_i$ and  can be ignored in the leading  approximation. Therefore, the fermionic decay channels  are one SM quark and one VLTQ. Using the Yukawa interactions in Eq.~(\ref{eq:hHQq}), the partial decay rates for $H\to (t u_J, b d_J)$ are given as
\begin{align}
\Gamma(H \rightarrow t_L \bar u_{JR} ) &= m_H \cos^2 \theta {\cal Y}^2_J \left( \frac{ (V_L^u)_{3J} }{4 \sqrt{\pi} } \right)^2 \left(1-r_t - r_J \right) \lambda^{\frac{1}{2}} \left( r_t, r_J \right)\,, \non \\
%
\Gamma(H \rightarrow b_L \bar d_{JR} ) &= m_H  \cos^2 \theta {\cal Y}^2_J  \left( \frac{ (V_L^d)_{3J} }{4 \sqrt{\pi} } \right)^2  \left(1- r_b - r_J \right) \lambda^{\frac{1}{2}} \left( r_b, r_J \right)
\label{eq:GHQq}
 \end{align}
with $r_f =m^2_f /m^2_H$ and  $\lambda(a,b)= 1+ a^2+b^2 -2a-2b -2ab$. By Eq.~(\ref{eq:mixing}), we see that the singlet $\Phi$ can couple to $W^+ W^-/ZZ$ through the mixing with SM Higgs. The decay rates for $H\to (W^+ W^-, ZZ)$ are found by 
\begin{align} 
\Gamma(H \rightarrow W^+ W^-) &=  \frac{m^3_H G_F}{8\sqrt{2} \pi} \sin^2\theta\left(1-4 r_W + 12 r^2_W \right) \sqrt{1-4 r_W}\,, \non \\
%
%
\Gamma(H \rightarrow Z Z) &=  \frac{m^3_H G_F}{16\sqrt{2} \pi} \sin^2\theta\left(1-4 r_Z + 12 r^2_Z \right) \sqrt{1-4 r_Z}\,.
%
\label{eq:GHVV}
\end{align}
The singlet $\Phi$ also couples to $hh$ through cubic term in Eq.~(\ref{eq:VHS}). The decay rate for $H \to hh$ is then obtained as
\be
\Gamma(H \rightarrow hh) = \frac{ G_F m_H (m_H^2-m_h^2)}{256 \sqrt{2} \pi }  \sin^2 2 \theta \sqrt{1 - 4 r_h}, \label{eq:GHhh}
\ed
 where  for simplicity we have used the  relation of $\sin \theta$ given below Eq.~(\ref{eq:Hmass}) by assuming $\lambda_3 =0$.
Additionally, by the effective coupling $ggH$ given in Eq.~(\ref{eq:LggH}), $H$ can also decay into gluon-pair and the corresponding partial decay width is found as
\be
\Gamma(H \rightarrow gg)  =m_H^3 \frac{ \alpha_s^2 }{32 \pi^3}  \cos^2\theta \left|\sum_{i=1,2}\frac{N_{F_i} y_i}{m^2_{F_i}}    F(\tau_i) \right|^2\,. 
\label{eq:GHgg}
\ed
Since the process $H\to \gamma\gamma$ depends on $\alpha_g=g^2/4\pi$ and is much smaller than $H\to gg$, here we neglect its contribution to the width of $H$. By summing up the partial decay widths shown in Eqs. (\ref{eq:GHQq}), (\ref{eq:GHVV}), (\ref{eq:GHhh}) and (\ref{eq:GHgg}), we plot the total width $\Gamma_H$ as a function of $\zeta$ and $\sin\theta$ in the left panel of Fig.~\ref{fig:Gamma}, where we have set $\zeta=\zeta_1=\zeta_2$ and used $(m_H, m_F)=(2, 1)$ TeV and $y=3$, and the NLO K-factor for $H\to gg$ is $K_{H\to gg}=1.35$ \cite{Djouadi:2005gi}. We find that $\Gamma_H < 100$ GeV occurs while $\sin\theta \leq 0.11$.  The branching ratio (BR) for $H\to W^+ W^- + ZZ$ is also shown in the right panel of Fig.~\ref{fig:Gamma}. We see that BR($H\to VV$) with $V=W,\, Z$ is sensitive to $\sin\theta$. Although  $H\to VV$ becomes dominant when $\sin\theta > 0.11$, however in this region we get $\Gamma_H \geq 100$ GeV. 
%

\begin{figure}[h!] 
\includegraphics[width=70mm]{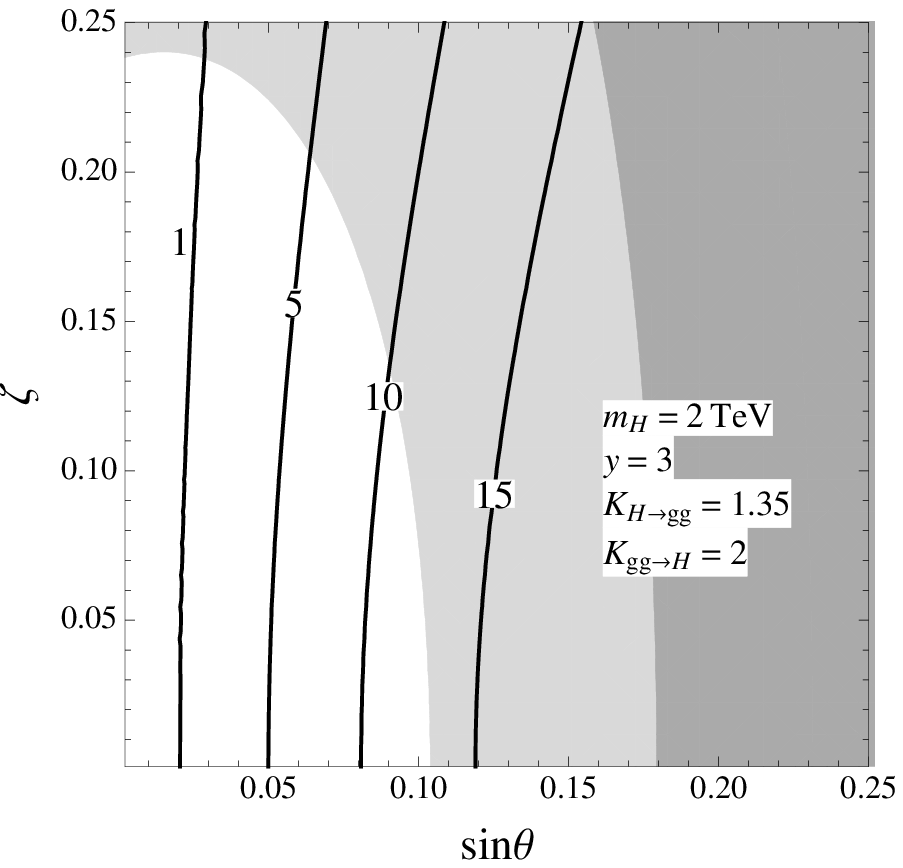} 
\includegraphics[width=70mm]{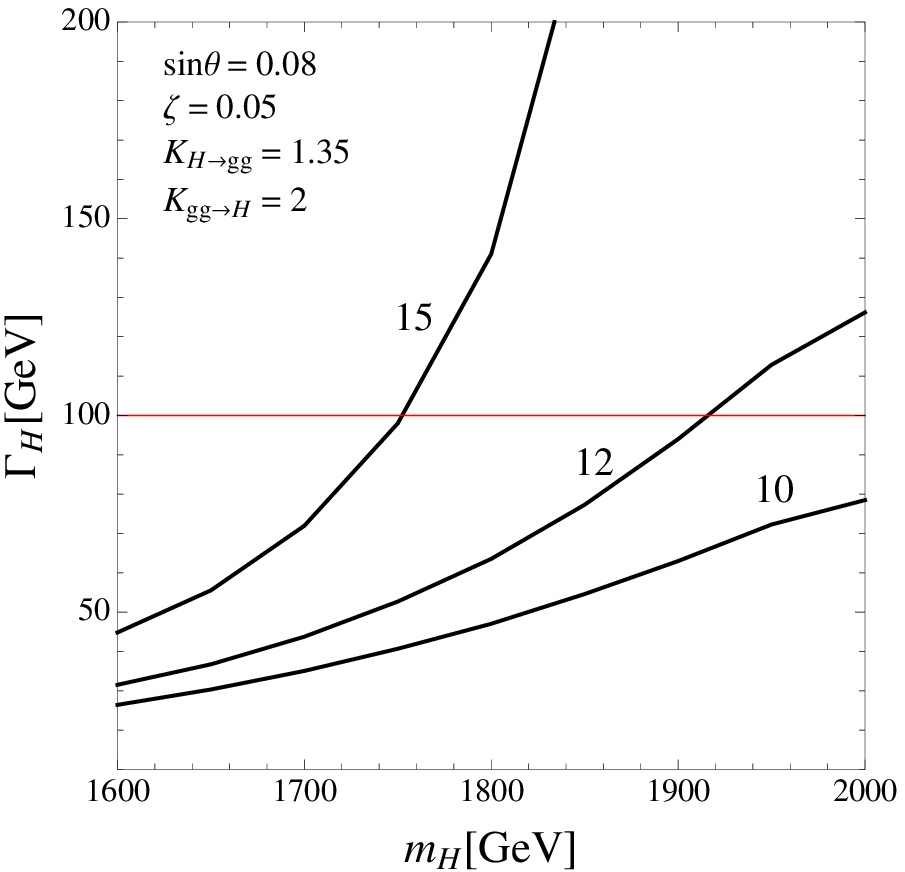} 
\caption{ Left:  contours for $\sigma(pp \to H) \times BR(H \to VV)$ (in units of fb) as a function of $\sin\theta$ and $\zeta$ with $(m_{H}, m_F)=(2, 1)$ TeV and $y=3$. Right: contours for $\sigma(pp \to H) \times BR(H \to VV)$ as a function of $m_H$ and $\Gamma_H$ with $\sin\theta=0.08$ and $\zeta=0.05$. }
\label{fig:XSBrH}
\end{figure}

Combing the results in Fig.~\ref{fig:XS} with those in Fig.~\ref{fig:Gamma}, we present the contours for $\sigma_{HVV}=\sigma(pp\to H)\times BR(H\to VV)$ as a function of $\zeta$ and $\sin\theta$ in the left panel of Fig.~\ref{fig:XSBrH} with $(m_H, m_F)=(2, 1)$ TeV and $y=3$. For $\sin\theta<0.11$, which leads to $\Gamma_H < 100$ GeV, we see that  $\sigma_{HVV}$ could still be of ${\cal O}(10)$ fb. Since the production cross section is very sensitive to $m_H$, for understanding the correlation between $\sigma_{HVV}$ and ($\Gamma_H$, $m_H$), we plot the contours for $\sigma_{HVV}$ as a function of $m_H$ and $\Gamma_H$ in the right panel of Fig.~\ref{fig:XSBrH}. We find that by reducing  $m_H$ of $15\%$, the value of $\sigma_{HVV}$ will be increased by $50\%$. 
%

\begin{figure}[h!] 
\includegraphics[width=70mm]{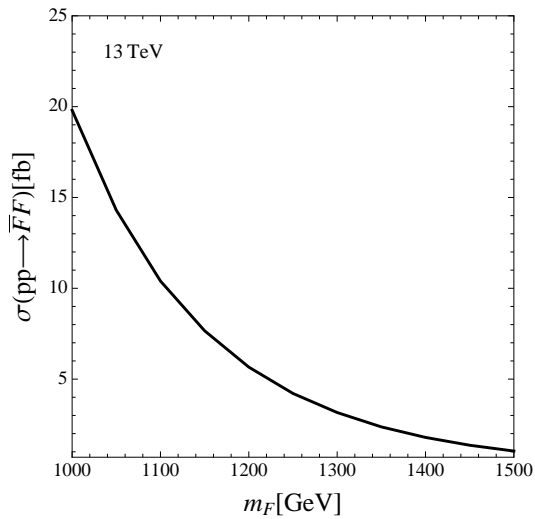} 
\caption{ Cross section for $pp\to F \bar F$  (in units of fb) as a function of $m_F$ at $\sqrt{s}=13$ TeV. }
\label{fig:XS13_VLQ}
\end{figure}

In the following we briefly discuss the implications of VLTQ model at the 13 TeV LHC. With the values of parameters that explain the diboson resonance, we find that the production cross section of  a single VLTQ is below 1 fb, therefore we study the production of VLTQ-pair and the possible decay channels. Due to QCD processes, the VLTQ-pair  production cross section is independent of the heavy quark flavor.  We present $\sigma(pp \to F \bar F)$  as a function of $m_F$ in Fig. \ref{fig:XS13_VLQ}, where $F$ stands for the possible VLTQ, i.e. $u_{4,5}$ and $d_{4,5}$, and the center of mass energy is $\sqrt{s}=13$ TeV.  For $m_F=1$ TeV, we get 20 fb for $\sigma(pp\to F \bar F)$. 

For studying the decays of VLTQs, we focus on the leading effects. By the Yukawa interactions in Eq.~(\ref{eq:hHQq}),  we see that VLTQs could decay into third generation SM quarks and SM Higgs. The partial decay widths for $u_J(d_J) \rightarrow t h (bh)$ are formulated by 
\begin{align}
\Gamma(u_{RJ} \rightarrow t_L h) &=  \cos^2 \theta \frac{G_F m_F^3}{\sqrt{2}} \left( \frac{(V_L^u)_{3J}}{4 \sqrt{\pi}}  \right)^2 \lambda^{\frac{1}{2}}(s_t,s_h) \tilde \lambda^{\frac{1}{2}}(s_t,s_h)\,, \nonumber \\
\Gamma(d_{RJ} \rightarrow b_L h) &=  \cos^2 \theta \frac{G_F m_F^3}{\sqrt{2}} \left( \frac{(V_L^d)_{3J}}{4 \sqrt{\pi}}  \right)^2 \lambda^{\frac{1}{2}}(s_b,s_h) \tilde \lambda^{\frac{1}{2}}(s_b,s_h) \label{eq:Fqh}
\end{align}
with $s_f = m^2_f / m^2_F$ and $\tilde{\lambda}(a,b)=1+a^2+b^2+2a-2b-2ab$. Besides the $h$-emission FCNC processes, according to Eqs.~(\ref{eq:WQq}) and (\ref{eq:ZQq}), VLTQs also can decay into SM quarks associated with $Z$- and $W$-emission. Consequently, the partial decay widths are given by
\begin{align}
\Gamma(u_{LJ} \rightarrow t_L Z) &= \frac{(g c_J^u)^2}{16 \pi c_W^2} m_F \lambda^{\frac{1}{2}}(r_t,r_Z) \left[\tilde \lambda^{\frac{1}{2}}(r_t,r_Z) + \tilde \lambda^{\frac{1}{2}}(r_Z,r_t) \frac{m_D^2 -m_Z^2-m_t^2}{m_Z^2} \right]\,, \nonumber \\
\Gamma(d_{LJ} \rightarrow b_L Z) &= \frac{(g c_J^b)^2}{16 \pi c_W^2} m_F \lambda^{\frac{1}{2}}(r_b,r_Z) \left[ \tilde \lambda^{\frac{1}{2}}(r_b,r_Z) +\tilde \lambda^{\frac{1}{2}}(r_Z,r_b) \frac{m_D^2 -m_Z^2-m_b^2}{m_Z^2} \right]\,, \nonumber \\
\Gamma(u_{LJ} \rightarrow b_L W) &= \frac{(g  \kappa_J^u)^2}{32 \pi } m_F \lambda^{\frac{1}{2}}(r_b,r_W) \left[ \tilde \lambda^{\frac{1}{2}}(r_b,r_W) + \tilde \lambda^{\frac{1}{2}}(r_W,r_b) \frac{m_D^2 -m_W^2-m_b^2}{m_W^2} \right]\,, \nonumber \\
\Gamma(d_{LJ} \rightarrow t_L W) &= \frac{(g  \kappa_J^d)^2}{32 \pi } m_F \lambda^{\frac{1}{2}}(r_t,r_W) \left[\tilde \lambda^{\frac{1}{2}}(r_t,r_W) +\tilde \lambda^{\frac{1}{2}}(r_W,r_t) \frac{m_D^2 -m_W^2-m_t^2}{m_W^2} \right] \,, \label{eq:FqZW}
\end{align}
where $\kappa^u_4 = -\zeta_1/2$, $\kappa^u_5 = \sqrt{2} \zeta_2$, $\kappa^d_4 = 0 $ and $\kappa^d_5 = -3 \zeta_2/2$. We see that if we take $\zeta_1 = \zeta_2$ and fix the value of $m_F$, the BR for VLQ decay is also fixed and independent of the free parameters $\zeta_{1,2}$.  We show the product of VLTQ-pair production cross section and BR of each VLTQ decay, denoted by $\sigma(pp\to F\bar F)\cdot BR(F\to f1) \cdot BR(\bar F\to f2)$, in Table~\ref{tab:XSBR}, where $m_F=1$ TeV is used and $f_{1(2)}$ stands for the possible decay channel of $F(\bar F)$. 
By the table, we find that the processes with cross section being larger than 5 fb are $u_{4,5} \to b W^+$, $d_4\to b Z$, $d_4 (\bar d_4) \to b h[Z](\bar b Z[h])$ and $d_5 \to t W^-$. It will be interesting to further investigate the significance of finding the VLTQ at the 13 TeV LHC. Since the detailed event simulation is beyond the scope of this paper, we leave the analysis for future work. 
   \begin{table}[h!]
   \caption{ Product of VLTQ-pair production cross section and branching ratio of VLTQ decay (in units of fb) denoted by $\sigma(pp\to F\bar F)\cdot BR(F\to f_1) \cdot BR(\bar F \to f_2)$, where $F(\bar F)$ is the VLTQ with electric charge of $2/3(-2/3)$ or $-1/3(1/3)$, $f_1 (f_2)$ is the decay channel listed in Eqs.~(\ref{eq:Fqh}) and (\ref{eq:FqZW})  and $m_F = 1000$ GeV is used.}
  \label{tab:XSBR}
 \begin{ruledtabular}
  \begin{tabular}{l|cccccc} 
$F\bar F$ & \quad $t \bar t hh$ \quad & \quad $t \bar t Z Z$ \quad & \quad $b \bar b W^+ W^-$ \quad & \quad $t \bar t h Z$ \quad & \quad $t \bar b W^- h (\bar t  b W^+ h) $ \quad & \quad $t \bar b W^- Z (\bar t  b W^+ Z) \quad $ \\ \hline
$u_4 \bar u_4$ &   $0.411$ & $1.44$ & $6.94$ & $1.53$ & $1.68$ & $3.16$ \\  
$u_5 \bar u_5$ &   $0.053$ & $0.188$ & $14.5$ & $0.200$ & $0.877$ & $1.65$ \\ \hline \hline
$F\bar F$ & \quad $b \bar b hh$ \quad & \quad $b \bar b Z Z$ \quad & \quad $t \bar t W^+ W^-$ \quad & \quad $b \bar b h Z$ \quad & \quad $b \bar t W^- h (\bar b  t W^+ h) $ \quad & \quad $b \bar t W^- Z (\bar b  t W^+ Z) \quad $ \\ \hline
$d_4 \bar d_4$ &   $2.13$ & $9.01$ & $0$ & $8.79$ & $0$ & $0$ \\  
$d_5 \bar d_5$ &   $0.015$ & $0.063$ & $16.8$ & $0.061$ & $0.497$ & $1.03$ \\  
  \end{tabular}
  \end{ruledtabular}
\end{table}

In summary, a  diboson resonance with mass of $1.8-2$ TeV is indicated by the observations of ATLAS and CMS. We study the possibility of Higgs singlet to explain the diboson excess. In order to enhance the heavy Higgs production cross section and to reduce the width to be ${\cal O}(100)$ GeV, we extend the standard model to include two vector-like triplet quarks and a Higgs singlet.  For escaping the constraints from the current SM Higgs measurements, we also include a color- and $SU(2)_L$-triplet scalar $(3,3)_{1/3}$. As a result, $pp\to h$ could be the same as that in the SM and $BR(h\to \gamma\gamma) \approx 0.93 BR_{\rm SM}(h\to \gamma\gamma)$. The result is consistent with current LHC data.  With six new heavy vector-like quarks and masses of 1 TeV, we find that the 2 TeV Higgs production cross section could reach ${\cal O}(20)$ fb if the Yukawa couplings are around 3, and its total width could be below ${\cal O}(100)$ GeV if the mixing angle between singlet and standard model Higgs  is $\sin\theta \leq 0.11$. The heavy Higgs production cross section is very sensitive to its mass, in order to understanding the mass dependence,  the dependence is shown in the right panel of Fig.~\ref{fig:XSBrH}. For discussing the implications of VLTQs at the LHC, we calculate $\sigma(pp\to F \bar F)$ at 13 TeV and find that with $m_F=1$ TeV, the production cross section is $20$ fb. Additionally, we also study the multiplication of  $\sigma(pp\to F \bar F)$ and the BRs of $F$ and $\bar F$ decays. We find that the cross section in the decay channel, such as $u_{4,5} \to b W^+$,  $d_4 \to b h(Z)$ and $d_5 \to t W^-$, could be over 5 fb. The detailed event simulation will be investigated in the future work. \\

\noindent{\bf Acknowledgments}

 This work is partially supported by the Ministry of Science and Technology of 
R.O.C. under Grant \#: MOST-103-2112-M-006-004-MY3 (CHC). 

\end{document}